\documentclass{ws-rv9x6}
\usepackage{subfigure}   
\usepackage{ws-rv-thm}   
\usepackage{ws-rv-van}   
\makeindex

\begin{document}

\chapter[Baryons and Chiral Symmetry]{Baryons and Chiral Symmetry}\label{ra_ch1}

\author[Keh-Fei Liu]{Keh-Fei Liu \footnote{liu@pa.uky.edy}}

\address{Dept. of Physics and Astronomy, University of Kentucky\\
Lexington, KY 40506 USA}

\begin{abstract}
The relevance of chiral symmetry in baryons is highlighted in three
examples in the nucleon spectroscopy and structure. The first one is the 
importance of chiral dynamics in understanding the Roper resonance.
The second one is the role of chiral symmetry in the lattice calculation of 
$\pi N \sigma$ term and strangeness. The third one is the role of chiral $U(1)$ 
anomaly in the anomalous Ward identity in evaluating the quark spin and the 
quark orbital angular momentum. Finally, the chiral effective theory 
for baryons is discussed.
 \end{abstract}
\body


\section{Introduction}\label{ra_sec1}

It is well-known that for three-flavor Quantum Chromodynamics (QCD), the
chiral symmetry $SU(3)_L \times SU(3)_R$ is spontaneously broken to
the diagonal $SU(3)_V$ with the octet pseudoscalar mesons as the Goldstone boson. 
There is also a $U_A(1)$ anomaly which begets a heavy $\eta'$ meson. As such,  
pions and chiral symmetry are important for low-energy hadron physics and 
the chiral dynamics has been successfully applied in subjects 
such as  $\pi\pi$ scattering~\cite{wei66,gl85}, vector dominance~\cite{sak66},
KSRF relation \cite{ksr66}, low-energy $\pi N$ scatterings~\cite{wei66,tom66}, 
$\pi N$ scattering up to about 1 GeV with the skyrmion \cite{mat87}, 
nucleon static properties~\cite{anw83}, electromagnetic 
form factors~\cite{anw83}, $\pi NN$ form factor~\cite{ll87}, and 
the Goldberger-Treiman relation~\cite{gt58}. 

The long range part of the nucleon-nucleon potential is due to  one
pion exchange and more modern NN potential has included the correlated 
two-pion exchange potential to account for the intermediate range attraction~\cite{bj76,wl81}.
These are the major ingredients in the realistic phase-shift equivalent NN 
potentials~\cite{NNpotential}. 

Realization and application of chiral symmetry in hadrons and nuclei has been a major research 
theme in Gerry Brown's scientific career. These include chiral symmetry in nucelon-nucleon 
interaction~\cite{brown79}, meson exchange currents~\cite{rb72}, and his joint research 
effort with Mannque Rho on little chiral bag~\cite{br79}, Brown-Rho scaling~\cite{br91}, and
dense nuclear matter~\cite{Brown:2001nh}.

With the advent of Quantum Chromodynamics (QCD) and the lattice chiral fermion
formulations in domain wall fermion~\cite{Kaplan:1992bt} and overlap fermion~\cite{neu98},
lattice QCD has provided a new tool in addressing the role of chiral
symmetry and the associated dynamics in first principle's calculation in terms
of quarks and gluons.

In this memorial manuscript, I will discuss 3 examples in baryons where chiral symmetry 
plays a crucial role as a way to pay tribute to Gerry's teaching throughout 
the author's professional career and to echo his passion of chiral symmetry by extending
the study of chiral symmetry from nuclear structure to nucleon structure~\cite{Liu:2014wba}.

The 3 examples are the Roper resonance, the pion nucleon sigma term and strangeness,
and the quark spin as representative cases for the importance of chiral symmetry in baryon spectroscopy 
and structure. 

\section{Roper resonance}

  The Roper resonance has been studied extensively, but its status as the lowest
excited state of the nucleon with the same quantum numbers is intriguing.
First of all, it has been noted for a long time that it is
rather unusual to have the first positive parity excited state lower than the
negative parity excited state which is the $N_{1/2}^-(1535)$ in the $S_{11}\, \pi
N$ scattering channel.  This is contrary to the excitation pattern in the meson sectors
with either light or heavy quarks.  This parity reversal has been problematic 
for the quark models based on $SU(6)$ symmetry with
color-spin interaction between the quarks~\cite{cr00} which cannot accommodate
such a pattern.  Realistic potential calculations with linear and Coulomb
potentials~\cite{lw83} and the relativistic quark model~\cite{ci86} all predict
the Roper to be $\sim$ $100$ -- $200\,{\rm MeV}$ above the experimental value
with the negative parity state lying lower.  On the other hand, the pattern of
parity reversal was readily obtained in the chiral soliton model like the
Skyrme model via the small oscillation approximation to $\pi N$
scattering~\cite{mat87}.  Although the first calculation~\cite{lzb84} of the
original skyrmion gives rise to a breathing mode which is $\sim 200\,{\rm MeV}$
lower than the Roper resonance, it was shown later~\cite{km85} that the
introduction of the sixth order term, which is the zero range approximation for
the $\omega$ meson coupling, changes the compression modulus and produces a better
agreement with experiment for both the mass and width in $\pi N$ scattering.

Since the quark potential model is based on the $SU(6)$ symmetry with residual
color-spin interaction between the quarks; whereas, the chiral soliton model is
based on spontaneous broken chiral symmetry, their distinctly different
predictions on the ordering of the positive and negative parity excited states
may well be a reflection of different dynamics as a direct consequence of the
respective symmetry.  This possibility has prompted Glozman and Riska to suggest~\cite{Glozman:1995fu}
that the parity reversal in the excited nucleon and $\Delta$, in contrast to
that in the excited $\Lambda$ spectrum, is an indication that the inter-quark
interaction of the light quarks is mainly of the flavor-spin nature, which implies
Goldstone boson exchange, rather than
the color-spin nature due to the one-gluon exchange.  This suggestion is
supported in the lattice QCD study of  ``Valence QCD"~\cite{Liu:1998um} where one
finds that the hyperfine splitting between the nucleon and $\Delta$ and also
between $\rho$ and $\pi$ are largely decimated when the $Z$-graphs in the 
quark propagators are removed.  This is 
an indication that the color-magnetic interaction is not the primary source of
the inter-quark spin-spin interaction for light quarks. (The color-magnetic
part, being spatial in origin, should be unaffected by the truncation of $Z$-graphs in
Valence QCD, which only affects the time part.)  Yet, it is consistent with the
Goldstone-boson-exchange picture which requires $Z$-graphs and thus the
flavor-spin interaction.

The failure of the $SU(6)$ quark model to delineate the Roper and its
photo-production has prompted the speculation that the Roper resonance may be a
hybrid state with excited glue~\cite{bc83} or a $qqqq\bar{q}$ five quark
state~\cite{khk00}.  Thus, unraveling the nature of Roper resonance has direct
bearing on our understanding of the quark structure and chiral dynamics of
baryons. 

\begin{figure}[htb]  \label{Compare_Roper}
  \centering
  {\includegraphics[width=1.1\hsize,height=0.6\textwidth]{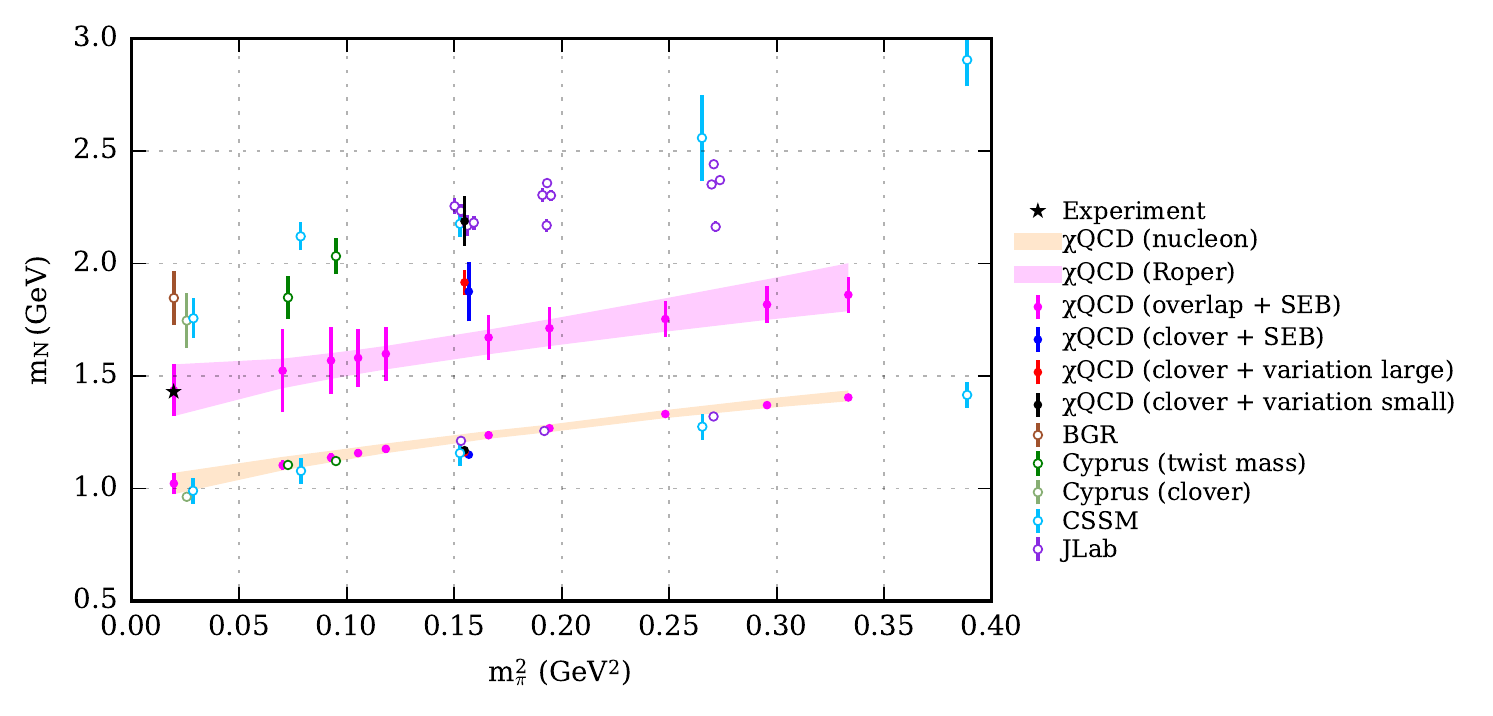}}
  \caption{Comparison of nucleon and Roper masses as a function of $m_{\pi}^2$ in several dynamical fermion calculations with different  fermion actions.}
\end{figure}

Lattice QCD is, in principle, the most desirable tool to adjudicate the
theoretical controversy surrounding the issue.  However, there is a complication.
As shown in Fig.~\ref{Compare_Roper}, the lattice calculations with the Clover 
fermions~\cite{edr11,mkl12,Alexandrou:2014mka}, chirally improved fermions~\cite{elm13}, and twisted mass
fermions~\cite{Alexandrou:2014mka} agree with the overlap fermion for the nucleon mass~\cite{Liu:2014jua}, but their 
first excited nucleon state (i.e. the Roper) are mostly over 2 GeV and much higher than those of the
overlap fermion in the pion mass range between 300 MeV and 600 MeV. Near the physical pion mass, they are
$\sim 300$ MeV above the experimental Roper mass at $\sim 1430$ MeV, while the overlap fermion
prediction agrees with experiment. 
This situation is basically a redux of the quenched calculations~\cite{mcd05,lhk07}, i.e. the overlap results
are much lower than those of the Wilson type fermions.
Why is there such a discrepancy? Since the calculations with Wilson type fermions, which breaks chiral symmetry, use the variational calculation,  while the calculation with the chiral overlap fermion  
adopts the Sequential Empirical Bayes Method (SEB)~\cite{Chen:2004gp} to extract the Roper state,
a question arises as to whether the discrepancy is due to different fitting algorithms or different actions at 
finite lattice spacing.

\begin{figure}[htb]  \label{Variation_smear}
\centering
  {\includegraphics[width=0.7\hsize,height=0.5\textwidth]{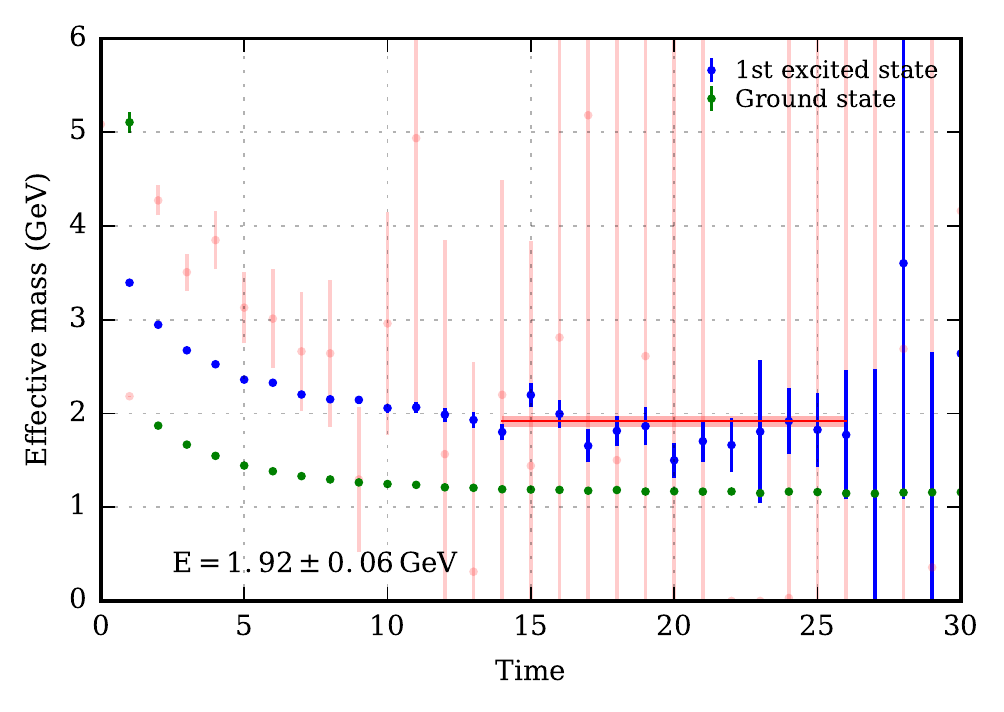}}
\centering
  {\includegraphics[width=0.7\hsize,height=0.5\textwidth]{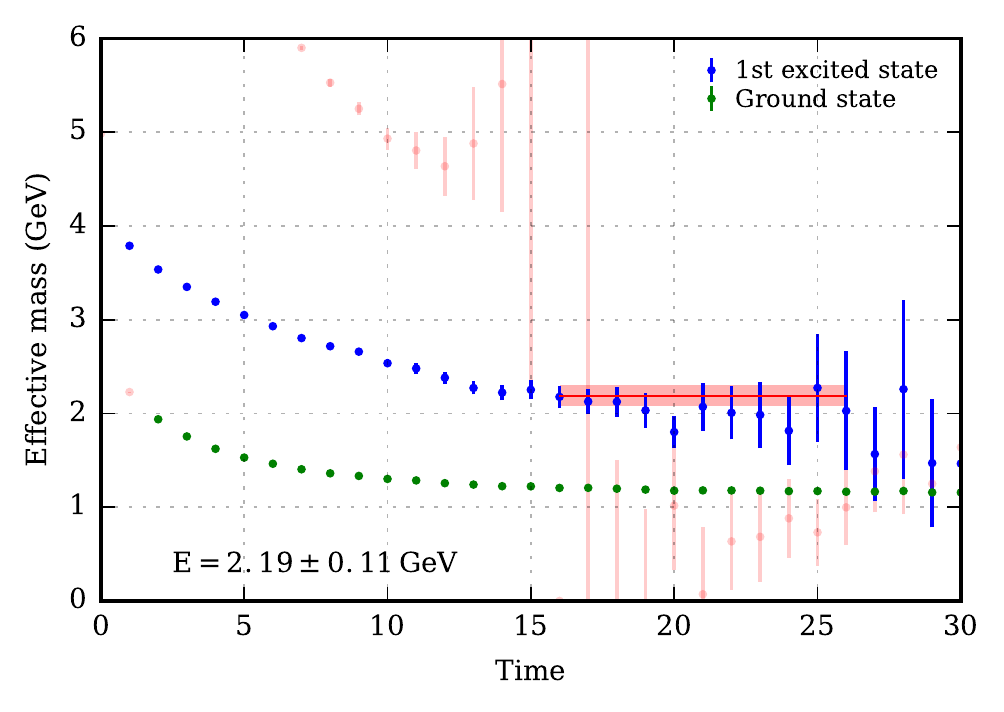}}
  \vspace*{-0.5cm}
  \caption{Nucleon and Roper masses from the variational method including large smeared sources with radius as
  large as 0.86 fm (upper panel) and with the radius of the smeared sources limited to less than 0.4 fm (lower panel).
}
\end{figure}

It was previously reported~\cite{Liu:2014jua} that to check the different algorithmic approaches, the SEB method
was used on the gauge configurations that are produced by HSC Collaboration~\cite{edr11} to calculate the nucleon 
and the Roper. These are $2+1$ flavor Clover fermion gauge configurations on the anisotropic 
$24^3 \times 128$ lattice with $a_s = 0.123$ fm and the light $u/d$ sea quark mass corresponds to a pion mass of 
$\sim 390$ MeV ({\it N.B.} the HSC results in Fig. 1 were obtained on the $16^3 \times 128$ lattice
with the same action and quark mass.) The SEB fitting of the nucleon and Roper masses are shown in Fig. 1 together with the variational results from HSC~\cite{edr11}. We see that while the nucleon mass agrees with that from 
HSC; the Roper, on the other hand, is lower 
than that from HSC by $\sim 300$ MeV (blue point in Fig. 1) with a $\sim 3 \sigma$ difference. 
To verify that this difference is not due to the fitting algorithm, we have carried out a variational calculation with different smearing sizes for the interpolation field. We see that when the r.m.s. radii of the Gaussian smeared source include one as large as 0.86 fm,  the Roper does appear lower at $1.92 (6)$ GeV as shown by the second plateau in Fig.~\ref{Variation_smear}
(upper panel) and indicated by the red point in Fig. 1. This is in agreement with the SEB result which
is shown as the blue point in Fig. 1. On the other hand,
when the r.m.s. radii of all the Gaussian smeared sources are less than 0.4 fm, the nucleon excited state is higher --
$2.19(11)$ GeV as shown in Fig.~\ref{Variation_smear} (lower panel) and indicated by the black dot in Fig. 1,
which is consistent with the HSC result indicated by the purple points in the same figure. This presumably confirms the 
speculation~\cite{Liu:2014jua} that one needs a large enough source to have a better overlap with the 2S excited state which has a radial node at $\sim 0.9$ fm from the study of its Coulomb 
wavefunction~\cite{Liu:2014jua}; whereas, smaller sources may couple to the 3S state strongly and results in a higher
mass. More importantly, the fact that this new variational result agrees with that from the SEB method  suggests that the SEB method
is a legitimate approach and, consequently, its extraction of the Roper at \mbox{$\sim 1.6$ GeV} in this pion mass range from the overlap fermion calculation in Fig. 1 should be reliable. 

The fitting algorithm issue being settled, this leads to the possibility that the difference is due to the different fermion
actions at finite lattice spacing. 

An extensive model has been constructed to
study the  $N^*$ resonance in $\pi N$ scattering partial waves~\cite{JuliaDiaz:2009ww}. The 
unperturbed states are the bare $N$ and $\Delta$ and the meson-baryon reaction channels including $\pi N, \eta N,$ and $\pi\pi N$ which has $\pi \Delta, \rho N$, and $\sigma N$ resonant components. This model fits the $\pi N$ scattering data well in various channels below 2 GeV. It is found~\cite{Suzuki:2009nj} in the $P_{11}$ channel, the meson-baryon transition amplitude is strong, which shifts the bare $1/2^+\, N^*$ at 1763 MeV down to ${\rm (Re M_R, -Im M_R)}$ = (1357, 76) MeV which corresponds to the $P_{11}$ pole of $N^*(1440)$. This is a shift of $\sim 400$ MeV in mass due to the meson-baryon coupling. 
Compared to the overlap fermion which has the lattice chiral symmetry which can have a larger  
$\langle 0|\chi_{N, 3q}|\pi N\rangle$ matrix element and a large Roper to $\pi N$ coupling. These 3 quark to 5 quark coupling which
invokes a pair creation or the Z-graph might be curtailed in the Wilson fermions like in ``Valence QCD'' as we discussed above and consequently results in a higher Roper state. This is a plausible explanation which is verifiable with
variational calculations of the Roper state with the Wilson-Clover fermion at smaller lattice spacings where chiral symmetry 
is better recovered.
Given the pattern of level reversal, the $\pi N$ scattering model and the lattice calculations,  we believe the Roper resonance has
a sizable $\pi N$ component in its wavefunction~\cite{Liu:2014jua} and is a showcase for the role of chiral dynamics.

\section{$\pi N$ and strangeness sigma terms}

As measures of explicit  and spontaneous chiral symmetry breaking in the baryon sector, $\sigma_{\pi N}$, defined as
\begin{eqnarray}
\sigma_{\pi N}\equiv \hat{m}\langle N|\bar{u}u+\bar{d}d|N\rangle,
\end{eqnarray}
where $\hat{m}=(m_u+m_d)/2$ is the averaged light quark mass, and $f_s^N$ defined as the strangeness $\sigma$ term 
as a faction of the nucleon mass 
\begin{eqnarray}
\sigma_{sN} \equiv m_s\langle N|\bar{s}s|N\rangle,  \,\,\,\,\,\,\, 
f_s^N = \frac{\sigma_{sN}}{m_N},
\end{eqnarray}
are fundamental quantities  which pertain to a wide range of issues in hadron physics. They include the quark mass contribution in the baryon which is related to the Higgs contribution to the observable matter~\cite{Young:2009zb,Bali:2016lvx}, the pattern of SU(3) 
breaking~\cite{Young:2009zb}, $\pi N$ and $KN$ scatterings~\cite{Brown:1971pn,Cheng:1972pe}, and kaon condensate in dense matter~\cite{Kaplan:1986yq}. Using the sum rule of  the nucleon mass, the heavy quark mass contribution can be deduced by that from the light favors, in the heavy quark limit and also in the leading order of the coupling~\cite{Shifman:1978zn,Gong:2013vja,Bali:2016lvx}. At the same time, precise values of the quark mass term for various flavors, from light to heavy, are of high interest for dark matter searches~\cite{Falk:1998xj,Ellis:2008hf,Giedt:2009mr}, where the popular candidate of dark matter (likes the weakly interacting mass particle) interacts with the observable world throughout the Higgs couplings, so that the precise determination of the $\sigma_{\pi N}$ and $\sigma_{sN}$ can provide constraints on the dark matter candidates.

Phenomenologically,  the $\sigma_{\pi N}$ term is typically extracted from the $\pi N$ scattering amplitude. To lowest order in $m_{\pi}^2$, 
the unphysical on-shell isospin-even $\pi N$ scattering amplitude at the Cheng-Dashen point corresponds to 
$\sigma(q^2=2m_{\pi}^2)$~\cite{Brown:1971pn,Cheng:1972pe} which can be determined from $\pi N$ scattering via fixed-$q^2$ dispersion 
relation~\cite{Cheng:1972pe}. $\sigma_{\pi N}$ at $q^2 = 0$ can be extracted through a soft correlated two-pion form 
factor~\cite{Gasser:1990ce,Becher:1999he,Pavan:2001wz}. Analysis of the $\pi N$ scattering amplitude to obtain $\sigma_{\pi N}(0)$ from the Lorentz covariant baryon chiral perturbation and the Cheng-Dashen low-energy theorem are also 
developed~\cite{Alarcon:2011zs,Chen:2012nx,Hoferichter:2015dsa}.  They give $\sigma_{\pi N}$ values in the range
$\sim 45 - 64$ MeV, while the most recent analysis~\cite{Hoferichter:2015dsa} gives 59.1(3.5) MeV.

Lattice calculations should be a good tool in giving reliable results to these quantities. Again, there is an issue
about chiral symmetry. It was pointed out~\cite{Michael:2001bv,Takeda:2010cw} that due to explicit chiral symmetry breaking, the quark mass in the 
Wilson type fermions has an additive renormalization and the flavor-singlet and non-singlet quark masses renormalize differently. 
In this case, the renormalized strange scalar matrix element $\langle N|\bar{s}s|N\rangle^R$ can be written as
\begin{equation}  \label{su3}
\langle N|\bar{s}s|N\rangle^R = \frac{1}{3} \Big[(Z_0 + 2 Z_8) \langle N|\bar{s}s|N\rangle + (Z_0 - Z_8)
 \langle N|\bar{u}u+ \bar{d}d|N\rangle\Big],
 \end{equation}
where $Z_0$ and $Z_8$ are the flavor-singlet and flavor-octet renormalization constants respectively.  $Z_0$ differs from $Z_8$ by a disconnected diagram which involves a quark loop. In the massless renormalization scheme, one can calculate
these renormalization constants perturbatively. For the massless case where 
$\bar{\psi}\psi = \bar{\psi}_L\psi_R +  \bar{\psi}_R\psi_L$,
a quark loop for the scalar density vanishes no matter how many gluon insertions there are on the loop, since
the coupling involving $\gamma_{\mu}$ does not change helicity. Thus, the massless scalar quark loop is zero and $Z_0 = Z_8$.
There is no mixing of the scalar matrix element with that of $u$ and $d$. This is the same with the overlap fermion, since
the overlap has chiral symmetry and the inverse of its massless quark propagator $D_c$ anti-commutes with $\gamma_5$, i.e. 
$\{D_c, \gamma_5\} = 0$ as in the continuum. 

\begin{figure*}
\centering
  \includegraphics*[scale=0.7]{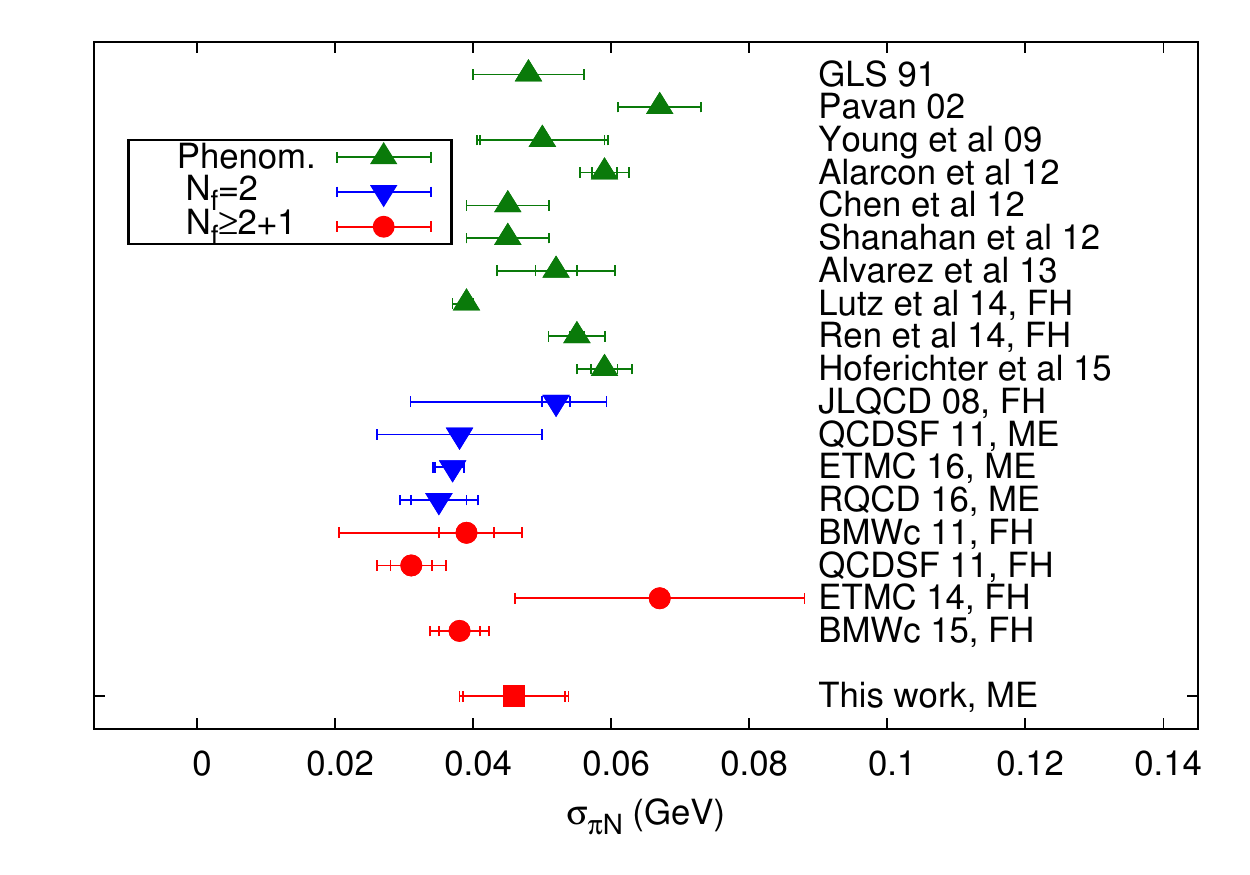} 
  \includegraphics*[scale=0.7]{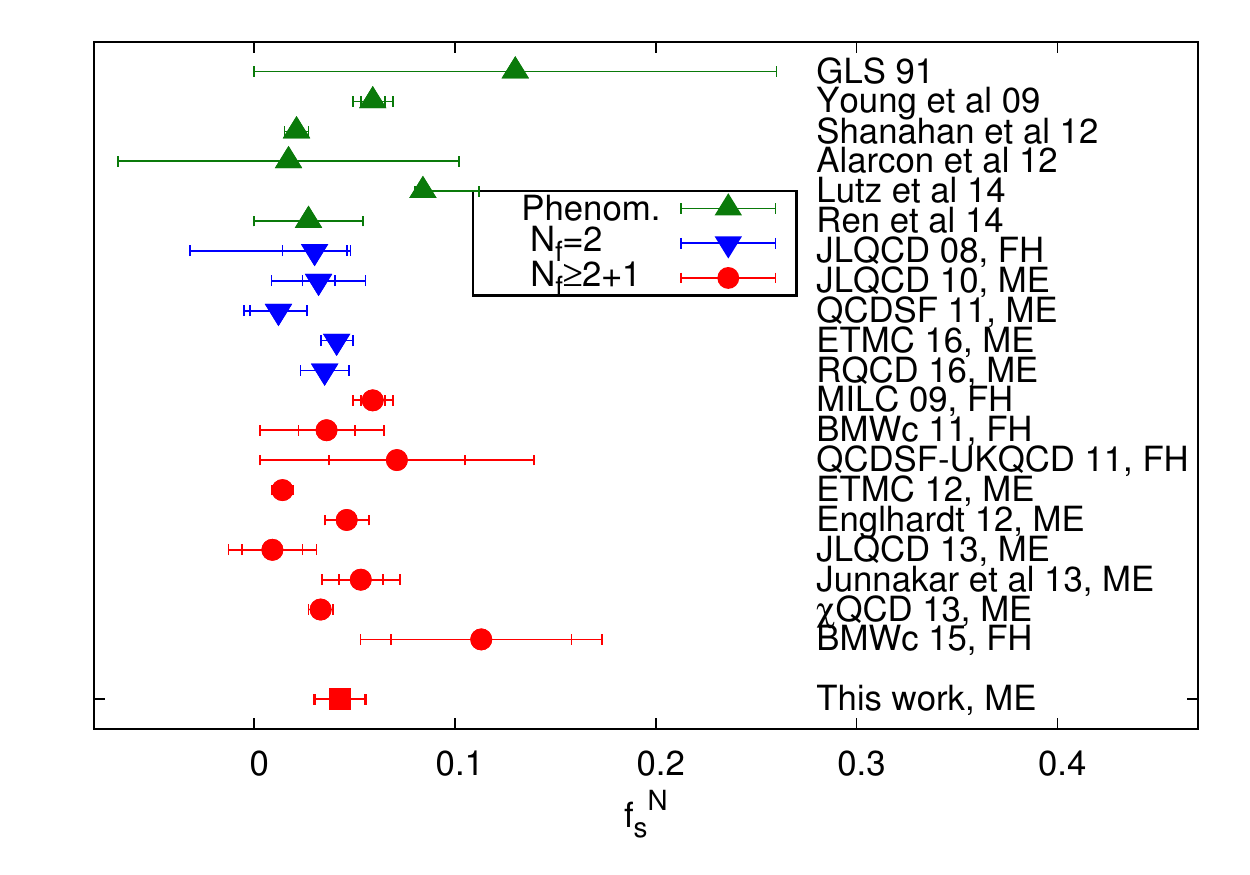} 
 \caption{The results of $\sigma_{\pi N}$ (upper panel) and $f^N_s$ (lower panel) from both phenomenology and lattice simulations. 
 The narrow error bar for each data point is the statistical, and the  broad one is that for the total uncertainty. The physical proton mass 938MeV is used to obtain $f^N_s$ in this work. They are color-coded in phenomenological and indirect approaches (green),
 $N_f =2$ lattice calculations (blue), and $N_f =2+1$ lattice calculations (red). 
 Detailed references  are given in Ref.~\cite{Yang:2015uis}.
  }\label{fig:hm2}
\end{figure*}

This is not so for Wilson type fermion where its
free quark propagator contains a term proportional to the Wilson $r$ term which violates chiral symmetry and will give a 
non-zero contribution to the scalar matrix element at the massless limit, leading to $Z_0 \neq Z_8$. Since the $u$ and $d$ matrix elements in the nucleon are not small, there can be a substantial flavor mixing at finite $a$. This lattice artifact 
due to non-chiral fermions can be removed by calculating $Z_0$ and $Z_8$~\cite{Bali:2011ks}. Furthermore, the  direct 
calculation of the matrix element with Wilson type fermions faces the complication that the sigma term with bare quark mass is not 
renormalization group invariant. This can also be corrected with the introduction of various renormalization constants to
satisfy the Ward identities~\cite{Bali:2011ks}. All of these involve additional work and will introduce additional errors. 
On the contrary,  there is no flavor mixing in the overlap fermion and the sigma terms are renormalization group invariant with bare mass and bare matrix element, since the renormalization constants of quark mass and scalar operator cancel, i.e. $Z_m Z_s =1$ due to
chiral symmetry. For the latest calculation with overlap fermion on $2+1$ flavor domain wall fermion gauge configurations for
several ensembles with different lattice spacings, volume, and sea masses including one at the physical pion mass, the global fit
gives the prediction of $\sigma_{\pi N} = 45.9(7.4)(2.8)$ MeV and $\sigma_{s N}= 40.2(11.7)(3.5)$ MeV. This
value of $\sigma_{\pi N}$ has a two-sigma tension with the recent results based on Roy-Steiner equations~\cite{Hoferichter:2015dsa} which gives $\sigma_{\pi N} = 59.1(3.5)$ MeV.

    To conclude, we believe that to calculate $\sigma_{\pi N}$ and $\sigma_{sN}$ which are fundamental quantities reflecting both the
explicit and spontaneous chiral symmetry, it is theoretically clean and straightforward procedure-wise to calculate them with
chiral fermions on the lattice in order to obtain reliable results without the complication of renormalization and flavor-mixing as compared to non-chiral fermions.

\section{Quark spin and orbital angular momentum}  \label{spin}

The quark spin content of the nucleon was found to be much smaller than that expected from the quark 
model by the polarized deep inelastic lepton-nucleon scattering experiments and the 
recent global analysis reveals that the total quark spin contributes only $\sim 25\%$ to the proton 
spin~\cite{deFlorian:2009vb}. This is dubbed `proton spin crisis' since no model seems to be able to
explain it convincingly and, moreover, quantitatively.

Once again, first principle lattice calculation should be able to address this issue. The ideal calculation would
be to use the conserved axial-vector current  of the chiral fermions which satisfies the anomalous Ward identity (AWI) on
lattice at finite lattice spacing. However, it is somewhat involved to construct the current itself for the overlap  
fermion~\cite{Hasenfratz:1998ri}. Before it is implemented, one can use the AWI as the normalization condition for
the simpler local axial-vector current
\begin{equation}\label{awi}
\partial_{\mu} \kappa_A  A_{\mu}^1 =    2m P  - 2i N_f  q ,
\end{equation}
where $A_{\mu}^1=  \sum_{i=u,d,s}\overline{\psi}_i i \gamma_{\mu}\gamma_5 (1-\frac{1}{2}D_{ov}) \psi_i$ is the 
local singlet axial-vector current and \mbox{$mP=\sum_{i=u,d,s}m_i \overline{\psi}_i i \gamma_5(1-\frac{1}{2}D_{ov}) \psi_i$} is the pseudoscalar density  with $D_{ov}$ being the massless
overlap operator and $q$ the local topological 
charge as derived in the Jacobian factor from the fermion determinant under the chiral transformation whose local version
is equal to $\frac{1}{16 \pi^2} tr_c G_{\mu\nu} \tilde{G}_{\mu\nu}(x)$ in the continuum~\cite{Kikukawa:1998pd}, i.e. 
\begin{eqnarray}  \label{top-charge}
\!\!q(x)\! =\!  {\rm Tr} \,\gamma_5 ( \frac{1}{2}D_{ov}(x,x) \!-\!1) 
 {}_{\stackrel{\longrightarrow}{a \rightarrow 0}} \frac{1}{16 \pi^2}tr_c \, G_{\mu\nu} 
\tilde{G}_{\mu\nu}(x).
\end{eqnarray}
$\kappa_A$ in Eq.~(\ref{awi}) is the finite lattice renormalization factor (often referred to as $Z_A$ in the literature
for the flavor non-singlet case) needed for the local axial-vector current to satisfy the AWI on the lattice with finite lattice spacing,
much like the finite renormalization for the vector and non-singlet axial-vector currents. We shall call it lattice normalization.
On the other hand, the $mP$ and $q$ defined with the overlap operators do not have multiplicative renormalization. There is
a two-loop renormalization of the singlet $A_{\mu}^1$ and the topological charge $q$ mixes with $\partial_{\mu} A_{\mu}^1$.
It turns out that they are the same. Thus, the renormalized AWI is the same as the unrenormalized AWI (but normalized) to
the $\alpha_s^2$ order. 
To utilize the AWI, one needs to calculate the matrix elements of $2mP$ and $2q$ on the r.h.s. of the AWI and extrapolate to
$q^2 =0$. However, the smallest $|q^2|$ is larger than the pion mass squared on the lattices that we work on, the extrapolation to
$q^2$ is not reliable. Instead, we shall  match the form factors at finite $|q^2|$ from both sides, i.e. 
\begin{equation}  \label{AWIFF}
2m_N  \kappa_A g_A^1 (q^2) + q^2 \kappa_A h_A^1(q^2) = 2m g_P (q^2) + N_f g_G(q^2).
\end{equation}
where the singlet $g_A^1 (q^2)$ and the induced pseodoscalar $h_A^1(q^2)$ are the bare form factors. $2m g_P(q^2)$ and
$g_G(q)$ are the form factors for the pseudoscalar current and topology respectively. 
From this normalization condition one can determine $\kappa_A$ and the normalized $g_A^1$ is $\kappa_A g_A^1(0)$.
This has been employed in the calculation of the strange quark spin to find 
$\Delta s + \Delta \bar{s}= - 0.0403(44)(61)$~\cite{Gong:2015iir}. This is  more negative than the other lattice calculations
with and axial-vector current, mainly because $\kappa_A = 1.36(4)$ is found to be larger than that of the flavor-octet axial-vector
current. The lesson here is that, unless the conserved current is used to carry out the calculation, it is essential to adopt the
AWI to obtain the normalization of the local axial-vector current. This is possible with the overlap fermion. 

While the final numbers on the $u$ and $d$ spin fraction which include the connected insertion are still beng worked out,
the initial results indicate that it is the larger negative $2mP$ matrix elements that cancel the positive topological charge term in the
triangle anomaly in the disconnected insertions that lead to a small $g_A^1$. 

There are various ways to decompose the proton spin into quark and glue spins and orbital angular 
momenta~\cite{Leader:2013jra,Liu:2015xha}. 
From the symmetrized energy-mometum tensor of QCD (the Belinfante form), it is shown~\cite{Ji:1996ek} that the proton
spin can be decomposed as
\begin{equation}       
\vec J_{\small{\rm QCD}} = \vec J_q + \vec J_g
  = \frac{1}{2} \vec\Sigma_q + \vec{L}_q + \vec{J}_g ,
\label{ang_op_def_split_2}
\end{equation}
where the quark angular momentum $\vec J_q$ is the sum of quark spin and orbital angular
momentum, 
\begin{equation}
  \vec{J}_q = \frac{1}{2} \vec\Sigma_q + \vec{L}_q  =  \int d^3x \, 
  \bigg{[} \frac{1}{2}\, \overline\psi\,\vec{\gamma}\,\gamma^5 \,\psi 
 + \psi^\dag \,\{ \vec{x} \times (i \vec{D}) \} \,\psi \bigg{]} ,
\label{quark_ang_op_split_1}
\end{equation} 
and each of which is gauge invariant. The glue angular momentum operator 
\begin{equation}
\vec{J}_g = \int d^3x \,\bigg{[} \vec{x} \times ( \vec{E} \times \vec{B} )\bigg{]} ,
\label{gluon_ang_op_def_split_1}
\end{equation}
is also gauge invariant. However, it cannot be further divided into the glue spin and orbital
angular momentum gauge invariantly with the Belinfante tensor.

Since it has a large finite volume effect to calculate the operator with a spatial $\vec{r}$ on the lattice with periodic boundary condition, one can instead calculate the quark and glue momentum and angular momentum from their form factors $T_1(q^2)$ and
$T_2(q^2)$ and obtain the momentum and angular momentum fractions from their forward limits, i.e. 
$\langle x\rangle = T_1(0)$ and $J = \frac{1}{2} (T_1(0) + T_2(0))$, much like the electric charge and magnetic moment
from the forward Dirac and Pauli form factors $F_1(0)$ and $F_2(0)$. After determining the quark angular momentum, 
the quark orbital angular momentum is obtained by subtracting the quark spin from it. This has been carried out in
a quenched approximation~\cite{Deka:2013zha}. The OAM fractions $2\langle L^q_\text{kin}\rangle$ for the $u$ and $d$ quarks in the CI have different signs and add up to $0.01(10)$, \emph{i.e.}
essentially zero. This is the same pattern which has been seen with dynamical fermion configurations and light quarks, as pointed out earlier. The large OAM fractions $2\langle L^q_\text{kin}\rangle$ for the $u$/$d$ and $s$ quarks in the DI is due to the 
fact that $g_A^1$ in the DI is large and negative, about $-0.12(1)$ for each of the three 
flavors. All together, the quark OAM constitutes a fraction of 
$0.47(13)$ of the nucleon spin. The majority of it comes from the DI. 

\begin{figure}[h]
  \centering
   \subfigure[] 
  {\rotatebox{0}%
    {\includegraphics[width=0.99\textwidth]{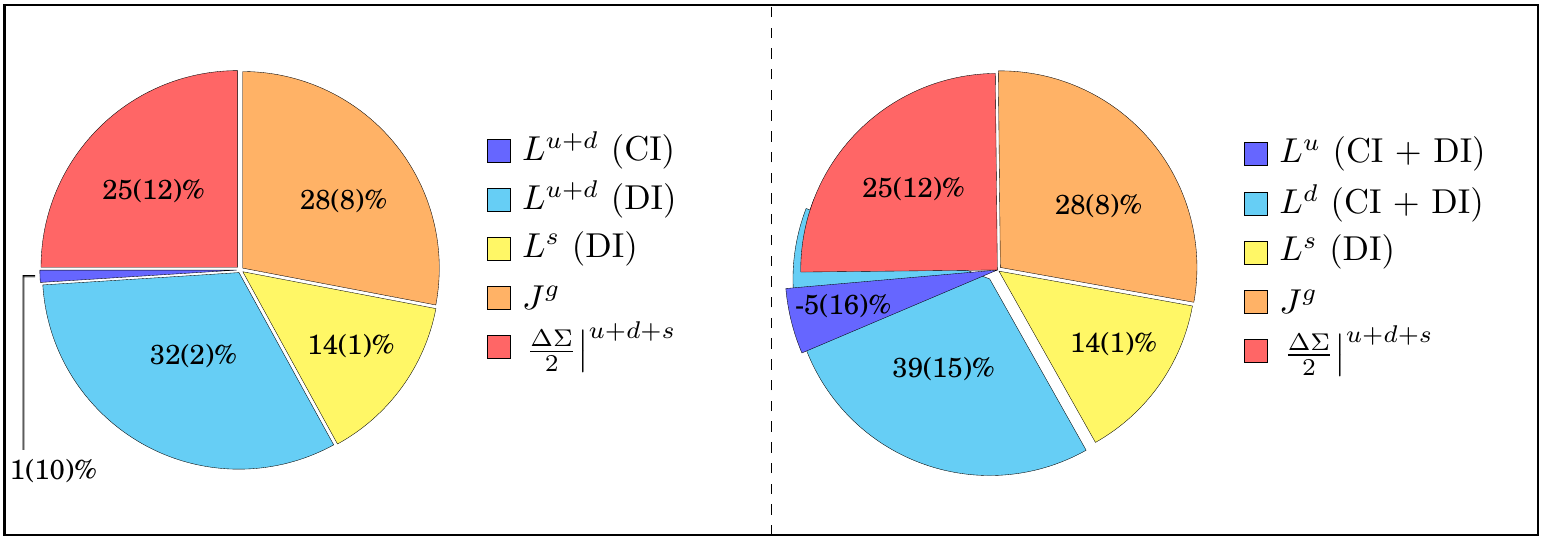}}
    \label{fig:pie_diag_orb_am}
  }
  \caption{Pie charts for the quark spin, quark orbital angular momentum and
  gluon angular momentum contributions to the proton spin.
        The left panel show the quark contributions separately for CI and DI,\ and the right panel 
    shows the quark contributions for each flavor with CI and DI summed together for $u$ and 
    $d$ quarks.%
  }
  \label{fig:pie_diag}
\end{figure}

As far as the spin decomposition is concerned, it is found that
the quark spin constitutes 25(12)\% of the proton spin, the gluon total AM takes 
28(8)\% and the rest is due to the quark kinetic OAM which is 47(13)\%. 

Since this calculation is based on a quenched approximation which is known to contain
uncontrolled systematic errors, it is essential to repeat this calculation with dynamical
fermions of light quarks and large physical volume. However, we expect that the quark OAM fraction may still be large in the dynamical calculation. 

In the naive constituent quark model, the proton spin comes entirely from the quark spin. On the 
other hand, in the Skyrme model~\cite{Adkins:1983ya} the proton spin originates solely from the OAM of the collective rotational motion of the pion field~\cite{Li:1994zp}. What is found in the present lattice calculation suggests that the QCD picture, aside from the gluon contribution, is somewhere in between 
these two models, indicating a large contribution of the quark OAM due to the meson cloud ($q\overline{q}$ pairs in the higher Fock space)  in the nucleon.

\section{Effective theory of baryons} \label{sec6}

Many estimates of quark spin and OAM contributions of the nucleon are based on quark models. However, quark models are not realistic effective theories of QCD, since they do not have chiral symmetry, a salient feature of QCD whose dynamics governs light-quark hadron structure, spectroscopy, and scattering at low energies. It is being learned quantitatively through lattice calculations of quark spin from the anomalous Ward identity~\cite{Gong:2015iir,Yang:2015xga,Liu:2015nva} that the smallness of the quark spin contribution in the nucleon is related to the $U(1)$ anomaly, the same anomaly which is responsible for the large $\eta'$ mass. 
This cannot be understood with quark models without the chiral $U(1)$ anomaly. Similarly, relativistic quark models do not explain the large OAM obtained from the lattice calculation in Sec.~\ref{spin}. Both the chiral quark model studies~\cite{Glozman:1995fu} and lattice calculation of valence QCD~\cite{Liu:1998um,Liu:2014jua} reveal that the level reversal of the positive and negative parity excited states of the nucleon, \emph{i.e.} $P_{11}(1440)$ (Roper resonance) and $S_{11}(1535)$, and the hyperfine splittings between the decuplet and octet baryons are dominated by the meson-mediated flavor-spin interaction, not the gluon-mediated color-spin interaction. 
All of these point to the importance of the meson degree of freedom  ($q\overline{q}$ pairs in the higher Fock space) which is missing in the quark model. 

\begin{figure}[t]
\begin{center}
\includegraphics[width=0.7\textwidth]{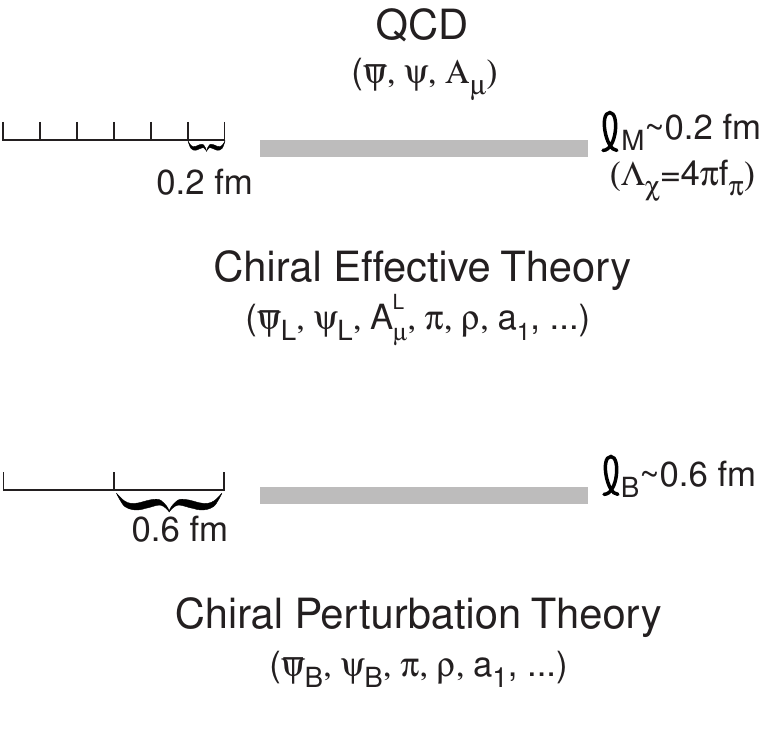}
\end{center}
\caption{A schematic illustration of the the two-scale delineation of
the effective theories. The shaded bars mark the positions of the 
cutoff scales $l_M$ and $l_B$  separating
different effective theories.}
\label{2scales}
\end{figure}

To see how this comes about, one can follow Wilson's renormalization group approach to effective theories. It is 
suggested by Liu {\it et al.}~\cite{Liu:1999kq} that the effective theory for baryons between the scale of 
$4\pi f_{\pi} (\sim 1 {\rm GeV})$, which is the scale of the meson size ($l_M\sim 0.2$ fm), and $\sim 300$ MeV, which is the
scale of a baryon size ($l_B\sim 0.6$ fm), should be a chiral quark model with renormalized couplings 
and renormalized meson, quark and gluon fields which preserve chiral symmetry. A schematic illustration
for such division of scales\footnote{We should point out that although
two scales are adopted here, they are distinct from those of
Manohar and Georgi~\cite{Manohar:1983md}. In the latter, the $\sigma$ -- quark model
does not make a distinction between the quark fields in the baryons and mesons. As such,
there is an ambiguity of double counting of mesons and
$q\overline{q}$ states. By making the quark-quark confinement length scale
$l_B$ larger than the quark-antiquark confinement length scale $l_M$, one
does not have this ambiguity.} for QCD effective theories is illustrated in Fig.~\ref{2scales}.

This suggestion is based on the observation that mesons and baryon form factor assume a monopole and dipole form, respectively. Since
the $\pi NN$ form factor is much softer than the $\rho \pi \pi$
form factor, it is suggested that the confinement scale of quarks in the
baryon $l_B$ is larger than $l_M$ -- the confinement scale between the 
quark and antiquark in the meson.
This is consistent with the large-$N_c$ approach to hadrons where the mesons 
are treated as point-like fields and the baryons emerge as solitons
with a size of order unity in $N_c$~\cite{Witten:1979kh}.
Taking $l_M$ from the $\rho \pi \pi$ form factor gives
$l_M \sim 0.2$ fm. This is very close to the chiral symmetry
breaking scale set by $\Lambda_{\chi} = 4 \pi f_{\pi}$. Considering
them to be the same, operators of 
low-lying meson fields become relevant operators below $\Lambda_{\chi}$. As for the baryon confinement
scale, Liu {\it et al.} take it to be the size characterizing the meson-baryon-baryon
form factors. Defining the latter from the respective meson poles in the nucleon pseudoscalar, 
vector, and axial-vector form factors in a lattice calculation~\cite{Liu:1998um} (see Fig. 17 
in the reference), they obtained $l_B \sim 0.6 - 0.7$ fm, satisfying $l_B> l_M$. Thus, in between these two scales $l_M$ and
$l_B$ (corresponding to the scale of $\sim 300$ MeV), one could have coexistence of mesons and quarks in an effective theory for baryons.

An outline is given~\cite{Liu:1999kq} to show how to construct a chiral 
effective theory for baryons. In the intermediate length scale between $l_M$ and $l_B$,
one needs to separate the fermion and gauge fields into long-range ones and
short-range ones
\begin{equation}
\psi = \psi_L + \psi_S, 
\qquad  A^{\mu} = A_L^{\mu} + A_S^{\mu},
\end{equation}
where $\psi_L/\psi_S$ and $A_L^{\mu}/A_S^{\mu}$ represent the infrared/ultraviolet 
part of the quark and gauge fields, respectively, with momentum 
components below/above $1/l_M$ or $\Lambda_{\chi}$. One adds irrelevant higher-dimensional operators to the
ordinary QCD Lagrangian with coupling between bilinear quark fields and auxiliary fields as
given in Ref.~\cite{Li:1995aw}, interpreting these quark fields 
as the short-range ones, \emph{i.e.} $\psi_S$ and $\overline{\psi}_S$.
Following the procedure by Li in Ref.~\cite{Li:1995aw}, one can integrate out the
 short-range fields and perform the derivative expansion to
bosonize $\psi_S$ and $\overline \psi_S$. This leads to
the Lagrangian with the following generic form:
\begin{equation}  \label{ect}
\begin{aligned}
{\cal L}_{\chi QCD} &= {\cal L}_{QCD'}(\overline{\psi}_L, \psi_L, A^{\mu}_L)
+ {\cal L}_M(\pi, \sigma, \rho, a_1, G,\cdots)  \\
&\quad+ {\cal L}_{\sigma q}
(\overline{\psi}_L, \psi_L, \pi, \sigma, \rho, a_1, G, \cdots). 
\end{aligned}
\end{equation}
${\cal L}_{QCD'}$ includes the original form of QCD but in terms of 
the quark fields $\overline{\psi}_L, \psi_L$, and
the long-range gauge field $ A^{\mu}_L$ with renormalized couplings. It 
also includes higher-order covariant derivatives~\cite{Warr:1986we}. 
${\cal L}_M$ is the meson effective
Lagrangian, \emph{e.g.} the one derived by Li~\cite{Li:1995aw} which should include the
glueball field $G$. Finally, ${\cal L}_{\sigma q}$ gives the coupling
between the $\overline{\psi}_L$, $\psi_L$, $G$ and mesons. As we see, in this
intermediate scale, the quarks, gluons and mesons coexist and
meson fields couple to the long-range quark fields.
Going further down below the baryon confinement scale $1/l_B$, one
can integrate out $\overline{\psi}_L$, $\psi_L$ and $A^{\mu}_L$, resulting in
an effective Lagrangian ${\cal L}(\overline{\Psi}_B, \Psi_B, \pi, \sigma, \rho, 
a_1, G,\cdots)$ in terms of the baryon and meson fields~\cite{Wang:1999xh}. This would 
correspond to an effective theory in the chiral perturbation theory.
In order for the chiral symmetry to be preserved, the effective theory of baryons at
the intermediate scale necessarily involves mesons in addition to the effective
quark and gluon fields. This naturally leads to a chiral quark effective theory. 

Models like the little bag model with skyrmion outside the MIT bag~\cite{Brown:1979ui}, the cloudy bag 
model~\cite{Thomas:1981vc} and quark chiral soliton model~\cite{Wakamatsu:2006dy} have
the right degrees of freedom and, thus, could 
possibly delineate the pattern of division of the proton spin with large 
quark OAM contribution. In particular, the fact that the $u$ and $d$ OAM in the MIT bag and to some extent the LFCQM in Table 1 have different signs from those of the lattice calculation may well be due to the lack of meson contributions as demanded by chiral symmetry in the effective theory of baryons.

\section{Summary}
    
    We discuss three examples in lattice QCD to highlight the role chiral symmetry
 and chiral dynamics play in baryons. From the observation of parity reversal of the excited the nucleon and $\Delta$ spectrum
 and the phenomenological model for the $\pi N$ scattering and $N^*$ states~\cite{}, it is suggested that it is the chiral dynamics
 that plays the leading role in the pattern of the low-lying baryon spectrum. This notion is supported by our study of the
 valence QCD and the fact that the lattice calculation of the Roper state by the non-chiral fermions at relatively coarse lattice spacing 
 are substantially higher  (by several hundred MeV's) than that of the chiral fermion which agrees with experiment.

 All the differences of lattice calculations with different fermion formulation are supposed to go away as the lattice spacing approaches zero due to universality. However, at finite lattice spacing, the different results form non-chiral and chiral fermions serve to
 illustrate the role of chiral symmetry and confirm the observation that chiral dynamic seems to be ubiquitous in low energy hadron physics. Thus, it is essential to incorporate  Goldstone bosons in the effective theory of baryons in addition to quarks and gluons below the chiral scale of  $4\pi f_{\pi}$.

\section{In memoriam}

  This manuscript is dedicated to the memory of Gerald E. Brown who was the author's Ph. D. thesis advisor, a mentor in his
 professional career and a lifelong friend. 
  
This work is is supported partially by US Department of Energy grant DE-SC0013065.


\end{document}